%% file: gorelov-dpf2009.tex
\documentclass[twocolumn,twoside,slac_two]{revtex4}
\usepackage{graphicx}
\usepackage{fancyhdr}
\input{ symblibrary.tex }

\begin{document}

  \title{ ATLAS Pixel Radiation Monitoring with HVPP4 System }
%

\author{Igor~Gorelov, Martin~Hoeferkamp, Sally~Seidel, Konstantin~Toms} 
\affiliation{Department of Physics and Astronomy, University of New
             Mexico, Albuquerque, NM 87131, USA}
\begin{abstract}
  In this talk we present the basis for the protocol for radiation
  monitoring of the ATLAS Pixel Sensors.  The monitoring is based on a
  current measurement system, HVPP4. The status on the ATLAS HVPP4
  system development is also presented.
\end{abstract}

\maketitle

\thispagestyle{fancy}


%
%
\input{ introduction/intro.tex }
%
%
\input{ current/curr.tex }
%
\input{ hardware/board.tex }
%
%
%
\input{ summary/summ.tex }
%
\end{document}

%% file: introduction/intro.tex
\section{Introduction}
\label{sec:intro}
  The High Voltage Patch Panel 4 (HVPP4) is a hardware system to connect
  and distribute and control the bias voltages to pixel sensors.  In
  this note we describe the extension of HVPP4 system to measure,
  digitize, and control the currents drawn by the pixel sensors
  comprising the ATLAS Pixel Detector. The HVPP4 current measurement
  system will be monitoring the pixel sensor currents {\it in situ} and
  in real time without requirement of special runs. The design topology
  of the system under development is discussed in technical
  note~\cite{hvpp4-tech-note}
\par 
  The ATLAS Pixel
  Detector 
  (see~\cite{Aad:2008zz}~\cite{jinst-mech-serv}~\cite{GarciaSciveres:2009zz}\cite{Golling:2009zz}~ 
  and a talk at this conference~\cite{Galyaev:2009fd}) comprises 1456
  pixel modules in the {\it Layer-0} (or {\it B-Layer}) , {\it Layer-1}
  and {\it Layer-2} from the barrel area, 288 modules mounted on 3 discs in
  the forward area and another 3 discs in the backward area. The total
  number of modules is 1744 units.  
%
%
  The geometry and module count for the barrel region of the pixel detector
  system are summarized in
  Table~\ref{table:barrel-parameters}. Modules are mounted on
  mechanical/cooling supports, called staves, in the barrel
  region. Thirteen modules are mounted on a stave and the stave layout
  is identical for all layers. The active length of each barrel stave
  is about 801 mm. The staves are mounted in half-shells manufactured
  from a carbon-fiber composite material. Two half-shells are joined
  to form each barrel layer.
\begin{table}[t]
  \begin{center}
  \caption{Basic parameters for the barrel region of the ATLAS pixel detector system.}
    \begin{tabular}{ |c|c|c|c|c|c| }
      \hline
      \multicolumn{1}{|c|}{Layer} & Mean &  Number of & Number of& Active \\
      Number & Radius [mm]  &  Modules & Channels & Area [m$^2$] \\
      \hline
      0 & 50.5  &  286 & 13,178,880 & 0.28 \\
      1 & 88.5  &  494 & 22,763,520 & 0.49 \\
      2 & 122.5 &  676 & 31,150,080 & 0.67 \\ 
      \hline
      \multicolumn{2}{|l|}{Total} & 1456 & 67,092,480 & 1.45 \\ 
     \hline
    \end{tabular}
    \label{table:barrel-parameters}
  \end{center}
\end{table}
  The two endcap regions are identical. Each is composed of three disk
  layers, and all disk layers are identical. The basic parameters of
  the endcap region are given in
  Table~\ref{table:endcap-parameters}. Modules are mounted on
  mechanical/cooling supports, called disk sectors. There are eight
  identical sectors in each disk.
\begin{table}[htbp]
  \begin{center}
  \caption{Basic parameters of the endcap region of the ATLAS pixel detector system.}
    \begin{tabular}{ |c|c|c|c|c|c|c| }
    \hline
     \multicolumn{1}{|l|}{Disk} & Mean $z$ & Number of & Number of& Active  \\
      Number & [mm] & Modules & Channels & {Area [m$^2$]} \\
      \hline
      0 & 495 & 48 & 2,211,840 & 0.0475 \\
      1 & 580 & 48 & 2,211,840 & 0.0475 \\
      2 & 650 & 48 & 2,211,840 & 0.0475 \\ 
      \hline
      \multicolumn{2}{|l|}{Total one endcap}   & 144 &  6,635,520 & 0.14 \\
      \multicolumn{2}{|l|}{Total both endcaps} & 288 & 13,271,040 & 0.28 \\ 
      \hline
    \end{tabular}
    \label{table:endcap-parameters}
  \end{center}
\end{table}
\par
  The pixel sensor consists of a $256 \pm3\,\mathrm{\mu m}$ thick
  n-bulk. The bulk contains n$^+$ implants on the read-out side and
  the p-n junction on the back side. For each sensor tile, the 47232
  pixel implants are arranged in 144 columns and 328 rows.  In 128
  columns (41984 or 88.9\,\%) pixels have implant sizes of
  $382.5\times 30\,\mathrm{\mu m}^2$ with a pitch corresponding to
  $400\times 50\,\mathrm{\mu m}^2$, and in 16 columns (5248 or
  11.1\,\%) pixels have implant sizes of $582.5\times 30\,\mathrm{\mu
  m}^2$ corresponding to a pitch of $600\times 50\,\mathrm{\mu m}^2$.
  In each column eight pairs of pixel implants, located near the
  center lines, are ganged to a common read-out, resulting in 320
  independent read-out rows or 46080 pixel read-out channels. This
  arrangement was chosen to allow for the connection of the sensor
  tile to 16 electronic front-end chips combined into a single 
  module.
\par 
  The sensitive area of $\sim$1.7\,$\mathrm{m^2}$ of the ATLAS pixel
  detector is covered with 1744 identical modules.
  Each module has an active surface of
  $6.08\times1.64\,\mathrm{cm^2}$.
\par 
  We assume that the dominant radiation damage type is displacement 
  defects in the
  bulk of the pixel sensor, caused by non-ionizing energy losses (NIEL).
  As the pixel barrel layers and disks are close to the interaction 
  point the charged pions dominate the bulk damage.
  These defects increase the reverse leakage current,
  degrade the charge collection efficiency, and change the effective
  doping concentration which directly determines the depletion
  voltage. The leakage current strongly depends upon the temperature
  of the pixel sensor and the particle fluence through the
  sensor volume. We define the fluence \flue as the number of
  particles causing damage equivalent to that of \( 1\mev \) neutrons
  traversing \( 1\cma \) of a sensor's surface.  The ATLAS pixel
  detector integrated fluence \( \flue \), (measured in \(
  {\mathrm{cm}}^{-2} \)), is expected to be proportional to integrated
  luminosity \( \IntL \), (measured in \( \invpb \)).
\par 
  The leakage current is monitored by the HVPP4 system at the pixel
  module granularity level.  The bias voltage to the sensors is provided
  by voltage channels of power supply modules from  
  Iseg~\cite{Iseg-specs}.
  During the first
  period of data taking, when the radiation damage of the sensors is
  small, 6 or 7 pixel modules will be fed by 1 Iseg power supply
  channel.  At some level of radiation damage after inversion of the
  sensors, before the current drawn by 6 or 7 pixel modules will reach
  the Iseg limit, a number of power supplies will be added until the
  system provides 1 Iseg power supply channel per pair of pixel modules.
\par 
  The current measurements on every module provide a powerful tool to
  monitor the status of every sensor, and hence the quality of the ATLAS
  Pixel Detector data.
%
%
  We will use the current measurements to estimate the fluence. We plan
  also to use the ad hoc ATLAS radiation monitoring
  devices~\cite{Hartert:2008zz} installed at several points of the pixel
  detector as well as the ones installed at other (outer) points of the
  ATLAS Inner Detector volume. That is, we will use two complementary
  data sets and methods to monitor the radiation in the ATLAS Pixel
  Detector physical volume.
\par 
  As the bias current depends on the sensor temperature, temperature
  measurements and related data are of crucial importance. We will use
  the sensor temperature data from the temperature probes with which
  every module is equipped.
%
%
%

%% file: current/curr.tex
\section{Leakage Current}
\label{sec:curr}
\par 
  The reverse bulk generation current's main cause is radiation
  damage of the crystal structure causing dislocations and other point
  defects.
\par
  Our analysis depends on the observation that increase in leakage
  current is proportional to fluence~\cite{Moll:1999nh},
  \begin{equation}
    \Delta{I} = \alpha\cdot\Phi_{eq}\cdot{V},
  \label{eq:i-phi}
  \end{equation}
  where \(\Delta{I} \) is the difference in leakage current at fluence
  \( \Phi_{eq} \) relative to the value before irradiation of the
  physical volume \( V \), and \( \alpha \) is the current-related
  damage coefficient. 
  The empirical parameter \(\alpha\) has been
  measured~\cite{Moll:1999nh} and found to be following:
  \begin{equation}
    \alpha(20\degc; 80\,min.\,@60\degc) = (3.99 \pm 0.03)\cdot{10^{-17}}\,\mathrm{A/cm},
  \label{eq:alpha-mich-moll}
  \end{equation}
  at 20\degc after annealing for 80 minutes at 60\degc.
\par 
  When considering the linear ansatz described above we must add
  the caveat that the ansatz is applied to the leakage currents drawn by
  sensors past their beneficial annealing time periods. We expect that
  at the beginning of data taking and during beneficial annealing
  periods the sensors will be drawing the currents at the low level of
  dark currents before the irradiation damage takes its course. This
  fact stipulates the need of a sensitivity to a lowest range of pixel
  sensor leakage currents when the HVPP4 system will be particularly
  useful as an excellent debugging tool.
  \begin{itemize}
    \item We analyze \( \Delta{I}/V \) (\( A/\cmc \)) vs \( \IntL \) 
          (\(\invpb \)) for each of the 1744 modules drawing current
          measured by the HVPP4 system.
    \item We assume that the fluence \flue is proportional to the
          integrated luminosity \IntL and the fitted slope of \(
          \Delta{I}/V \) vs \( \IntL \). Using the known \( \alpha \)
          and the slope we will infer the fluence \( \flue \) for each
          module.
    \item The current measurements are selected according to some
          quality criteria to be developed.
    \item The currents are corrected to a common temperature, 
          \( 20\degc \) (still to be decided).
  \end{itemize}
\subsection{Lifetime Estimate}
  By comparing current with integrated luminosity we assume that the
  linear fits of the temperature-corrected current readings per module
  can be extrapolated to predict the amount of current the pixel modules
  will draw after a certain integrated luminosity has been collected
  with the ATLAS pixel detector.
\par 
  Contrary to CDF SVX II, the ATLAS pixel \( S/N \) ratio is not an
  issue because the lowest noise level is determined by the sensor's
  design.
%
%
   However, leakage current in ATLAS Pixel Detector can lead to
   excessive power and thermal runaway, which basically limits the bias
   voltage that can be applied.
   {\bf A single Iseg power supply channel can sustain a maximum current
     of \( \mathbf{I\lsim{4000\mu{A}}} \)}~\cite{Iseg-specs}. Initially
   we have 6 or 7 pixel modules biased by a single Iseg power supply
   channel what gives us a maximum current to be reached in the range of
   \( I_{sensor}\lsim{(550...700)\mu{A}} \).
\par
  Another important predictor of a pixel sensor's lifetime is its
  depletion voltage.  We assume that the sensor will be kept biased at
  the full depletion voltage until limited by leakage current around 
  \( 550\,\mu{A} \) to \( 700\,\mu{A} \).  After that the sensor will
  operate partially depleted with reduced signal amplitude resulting in
  reduced hit efficiency.
\par
  The next two periods of a pixel sensor's life should be expected:
    \begin{itemize}
      \item[$\bullet$] The first years, operated at full depletion.  The
            end is determined by approaching
      \begin{itemize}
        \item[$\circ$] either a critical range of high currents with
            technically motivated cut-off values (we consider this case
            to be most probable),
        \item[$\circ$] or the maximum available bias voltage provided
              by the Iseg power supply at its channel level.
      \end{itemize}
      \item[$\bullet$] Later years of operation in partially depleted
            mode.  At this point the sensor draws high current, still
            within the safety margin or at the maximum available bias
            voltage, but its pixels' hit efficiencies gradually diminish
            with integrated luminosity (or absorbed fluence).
    \end{itemize}
\subsection{ATLAS Radiation Field Measurements}
  The radiation field inside the ATLAS Inner Tracker volume is measured
  by a number of standard ATLAS radiation monitors~\cite{Hartert:2008zz}.
  We are interested in the devices sensitive to hadron NIEL radiation rather
  than ionization as the expected bulk damage in pixel sensors comes
  from the ambient hadron (mostly pion) energy flow.
\par 
  The measurements will be processed and some model of the ATLAS
  radiation field will be developed. The measurements will be subjected to
  a fit by the model (similar to~\cite{D'Auria:2003qi}) with the requirements that
  \begin{itemize} 
%
    \item The radial dependence can be parametrized as a polynomial with
          an inverse powers terms included, e.g. as in Eq.~\ref{eq:flue-poly} and 
          the radial function can be fitted to ATLAS radiation monitors'
          data points
  \end{itemize}
\par
  {\it Layer-2} of the ATLAS pixel detector is equipped with standard
  ATLAS radiation probes. The \flue measured by radiation probes on 
  {\it Layer-2} should be compared with the results of fits of 
  \( \Delta{I}/V \)~(in~A/\cmc) vs \( \IntL \)~(in~\invpb) which are used to
  recalculate the \flue based on the known current related damage rate
  \( \alpha \) and the radiation to luminosity rate \( R_{dose} \)
  discussed in the subsection below.  The difference between the two
  measurements will give us an estimate of the systematic uncertainty of
  the method based on leakage currents.
\subsection{Expected Precision of the HVPP4 Current Measurements}
  We assume that the most critical aspect of the analysis 
  is the linear fit using Eq.~\ref{eq:i-phi}. Measurement statistics 
  will determine the fit errors of the slope parameter. 
  Therefore the predictions will involve 
  \begin{itemize}
    \item the HVPP4 precision on current measurements which should be taken as 
          a systematic uncertainty. 
          We expect that the precision \( \delta(\mathrm{HVPP4}) \) 
          of the current measurement board will be some fixed level of 
          current uncertainty determined by the circuitry of the board.
    \item the number of points, e.g. the number of data runs or {\it smaller 
          accessible data periods}  with corresponding 
          current measurements averaged over every data run or period.
    \item the uncertainty on the luminosity values provided by 
          the ATLAS luminosity group. This factor determines 
          the period defined by the ATLAS Central DAQ (e.g. run or run section)
           when the most reliable 
          \Lumi measurements made by the ATLAS luminosity monitors are available.
          After several years of data taking the uncertainty on \Lumi
          will reach \( \delta(\Lumi)\sim 6\% \) if it follows the  experience 
          of other experiments (H1, CDF, D\O). During the 
          first three years we expect the uncertainty to be larger, 
          \( \delta(\Lumi)\sim 10\% \). 
    \item Another contribution to the uncertainty is due to the error on \(\alpha\), 
          see Eq.~(\ref{eq:alpha-mich-moll}).
%
%
%
%
      \item CDF~\cite{Worm:2005wb}~\cite{cdf-steve-worm} used a more conservative estimate:
          \[ (3.0 \pm 0.6)\cdot{10^{-17}}\,\mathrm{A/cm\,,} \]
%
          From this we expect an uncertainty \( \delta(\alpha)\sim 20\% \).           
  \end{itemize}
%
%
\par  
  {\bf In conclusion, the HVPP4 current measurement precision 
  will be determined by some fixed uncertainty to be
  derived from engineering specifications. The uncertainty   
  should comply with dominating uncertainties 
  coming on \( \mathbf{\Lumi} \) and \(\mathbf{\alpha}\).} 
\subsection{ATLAS MC Simulation Results: Expected Flux and Fluence}
\label{subsec:mcsim}
  The radiation fields in the ATLAS Detector have been predicted with 
  a full MC simulation~\cite{atlas:rad-bgr}. The fluence dependence
  as a function of radius has been parametrized as in Eq.~\ref{eq:flue-poly},
 \begin{equation}
   \flue = ({a_{-2}}\cdot{r^{-2}} + {a_{-1}}\cdot{r^{-1}})/1000\invfb
 \label{eq:flue-poly}
 \end{equation}
 The polynomial coefficients are shown in Table~\ref{table:flue-poly}.
\begin{table}[htbp] 
  \begin{center}
  \caption{Fluence parametrization: the polynomial coefficients 
           for ATLAS Pixel Detector \( z\approx{0.0\cm} \)  
           position along the beam axis.}
  \begin{tabular}{|c|c|c|}
  \hline
  Mean \(z,\cm\) & \({a_{-2}}\) & \({a_{-1}}\) \\
  \hline
   \(0\)   & \(4.93\cdot10^{+16}\) & \(0.25\cdot10^{+16}\) \\
  \hline
  \multicolumn{3}{|c|}{ The numbers normalized to \( \IntL = 1000\invfb \) }\\
  \hline
  \end{tabular}
  \label{table:flue-poly}
  \end{center}
\end{table}
\par
  The model expressed by Eq.~\ref{eq:flue-poly} assumes that the MC simulation 
  results when the \(z\)-coordinate is set as \( z=0.0\cm \) is good for the 
  whole barrel region of the Pixel Detector. Moreover as the worst case scenario 
  for the pixel disk layers, the model is recommended~\cite{atlas:rad-bgr} 
  to be extended over the whole \(z\in( -650\mm, +650\mm )\) range 
  between end cap disk 3 layers. For \(z\in( -650\mm, +650\mm )\) the model 
  conjectures a cylindrical symmetry of the fluence field 
          \[ \Phi_{1MeV\,eq}\,(r,\phi,z)\equiv\flue\,(r) \]
  The next caveat should be added here: a possible LHC beam offset
  w.r.t. ATLAS geodetic center will break the cylindrical symmetry.
\par
\begin{figure}[htbp]
    \begin{center}
    \caption{ The fluence for \( \IntL = 10,\,100,\) of collisions,
      predicted in the Pixel Detector region for \( z=0.0\cm
      \)~\cite{atlas:rad-bgr}. The \(r\)-positions for {\it
        Layer-0,-1,-2} are shown with vertical
      lines. \label{fig:pix-fluence} }
      \includegraphics[width=85mm]{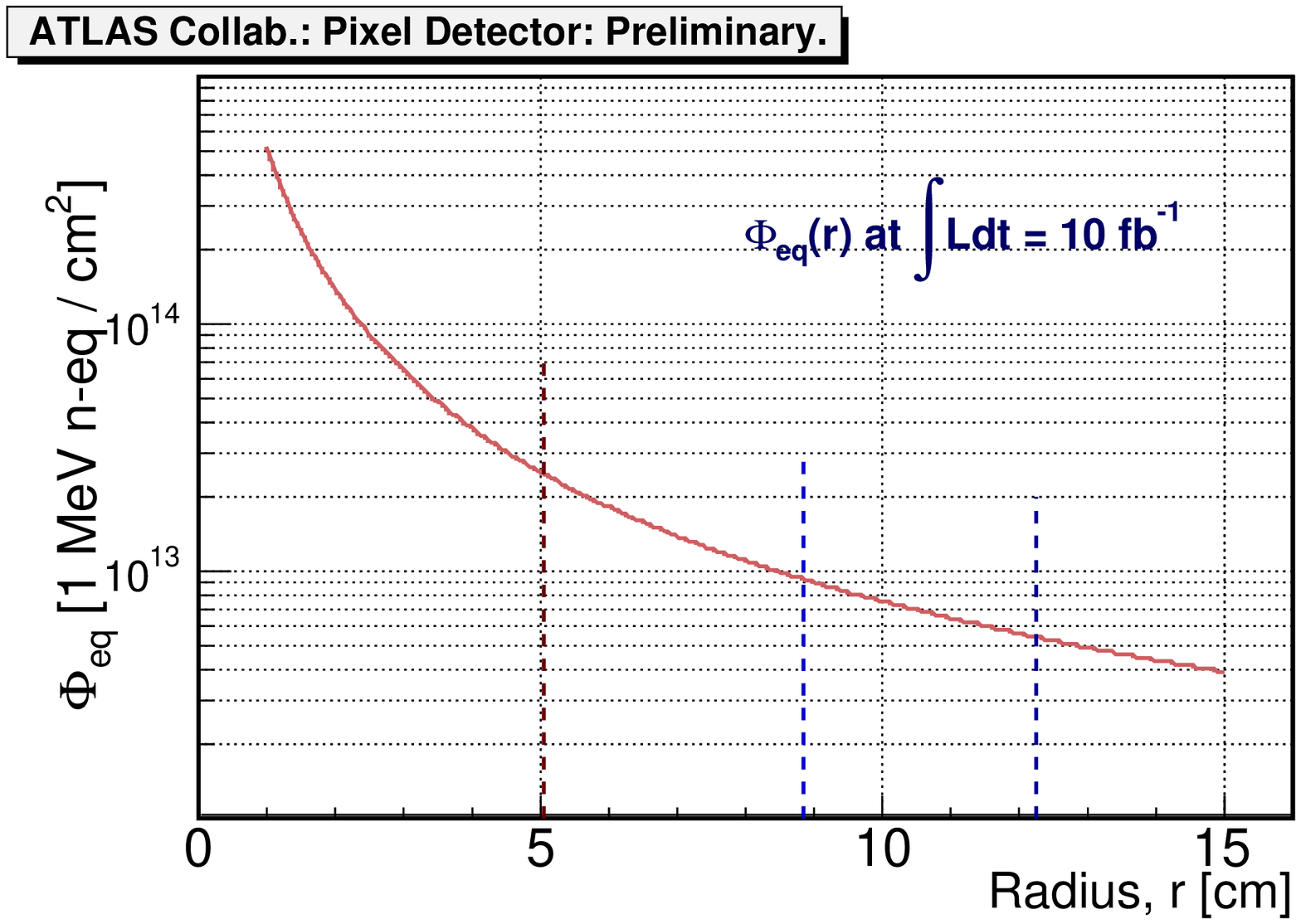}\\
      \includegraphics[width=85mm]{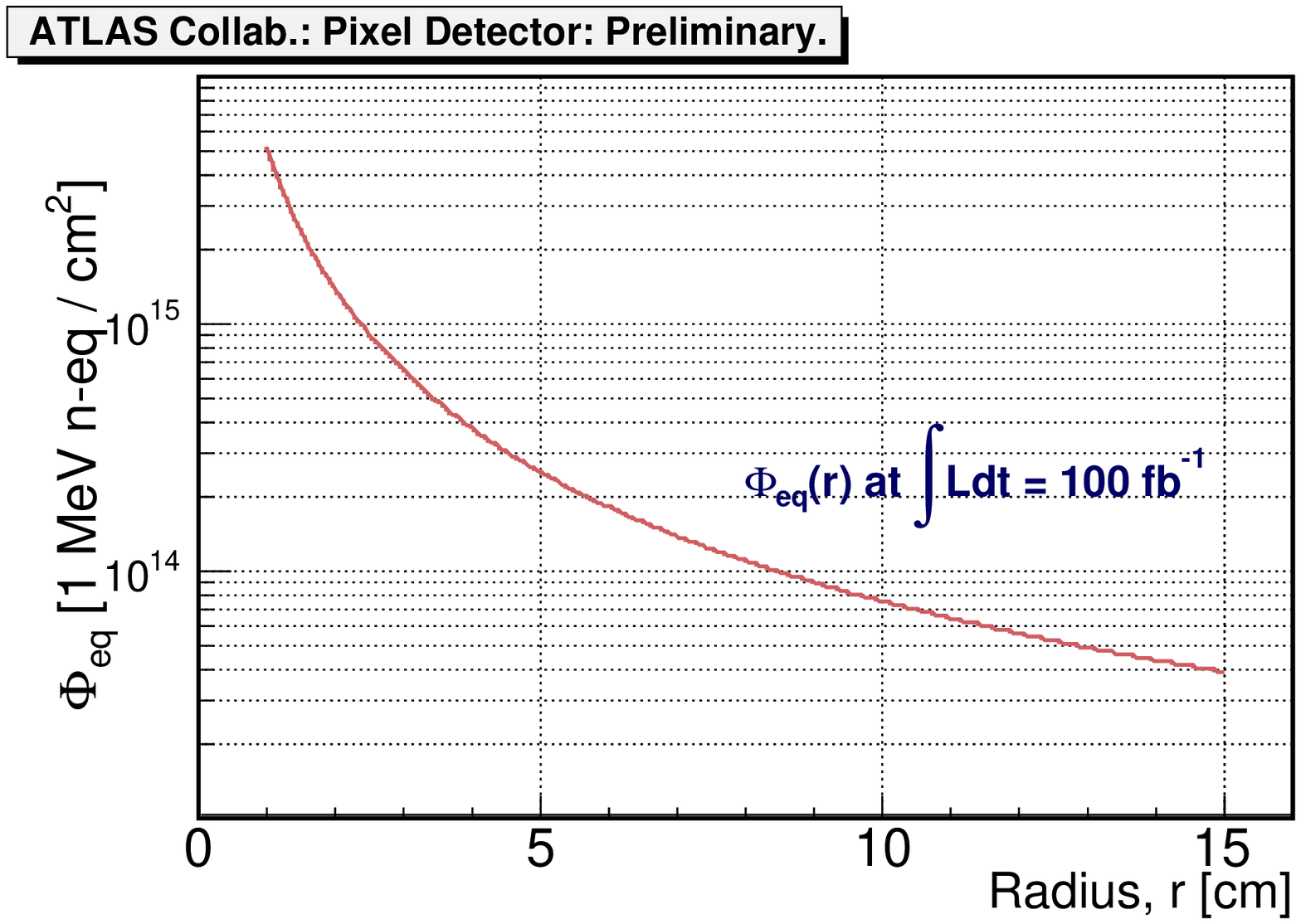}\\
    \end{center} 
  \end{figure}

 
  The fluence in the Pixel Detector area with \( z=0.0\cm \)
  simulation assumption~\cite{atlas:rad-bgr} for an integrated
  luminosity \( \IntL = 10\) and \(100\invfb \) is shown in
  Fig.~\ref{fig:pix-fluence}.
\subsection{Barrel Layers: Preliminary Estimates of the Slopes 
            of \( \Delta{I}/V \) versus \( \IntL \)}
  The slopes for the barrel area {\it Layer-0}, {\it Layer-1} and {\it Layer-2}
  are different: 
    \[ slope_{L0}\,>\,slope_{L1}\,>\,slope_{L2} \]
  We assume \( \alpha \) from Eq.~(\ref{eq:alpha-mich-moll}).  The
  slopes differ because the fluence depends strongly upon the radius
  as is shown in the previous subsection, \(\flue (r)\).  We will have
  experimental measurements from the standard ATLAS probes~\cite{Hartert:2008zz} 
  installed at 
  {\it Layer-2}, as well as the measurements to come from the other points
  instrumented by standard ATLAS radiation monitors~\cite{Hartert:2008zz}. 
   We can calibrate
  the parametrization model expressed by Eq.~(\ref{eq:flue-poly}) and
  shown in Table~\ref{table:flue-poly}, which was based on MC
  predictions using experimental points.
\par 
  Another conjecture we make here is that all modules on the
  same Iseg channel draw the same current. At real experimental 
  conditions the modules will be drawing different currents
  due to variations of temperature and other running conditions.
\par
\begin{figure}[htbp]
    \begin{center}
    \caption{ The expected currents \textbf{at maximum operational
              temperature values reached in the Fall 2008 Cosmic Run
              by the Pixel Detector} barrel {\it Layer-0}, 
              {\it  Layer-1} and {\it Layer-2} versus the integrated
              luminosity in two ranges: 
              \(\IntL\in(10\invpb,\,10\invfb) \) (upper plot) and 
              \(\IntL\in(10\invfb,\,1500\invfb) \) (lower plot).  The
              different slopes due to different fluences through 
              {\it Layer-0}, {\it Layer-1} and {\it Layer-2} are seen in
              \( \log \)-scaled coordinates as the offsets between
              lines. The higher operational temperature, \(7.1\degc\)
              for outer {\it Layer-2}, raises the current to values
              similar to those in {\it Layer-1} which is kept at
              temperature \(1.6\degc\). \label{fig:pix-maxtC-current}  } 
      \includegraphics[width=85mm]{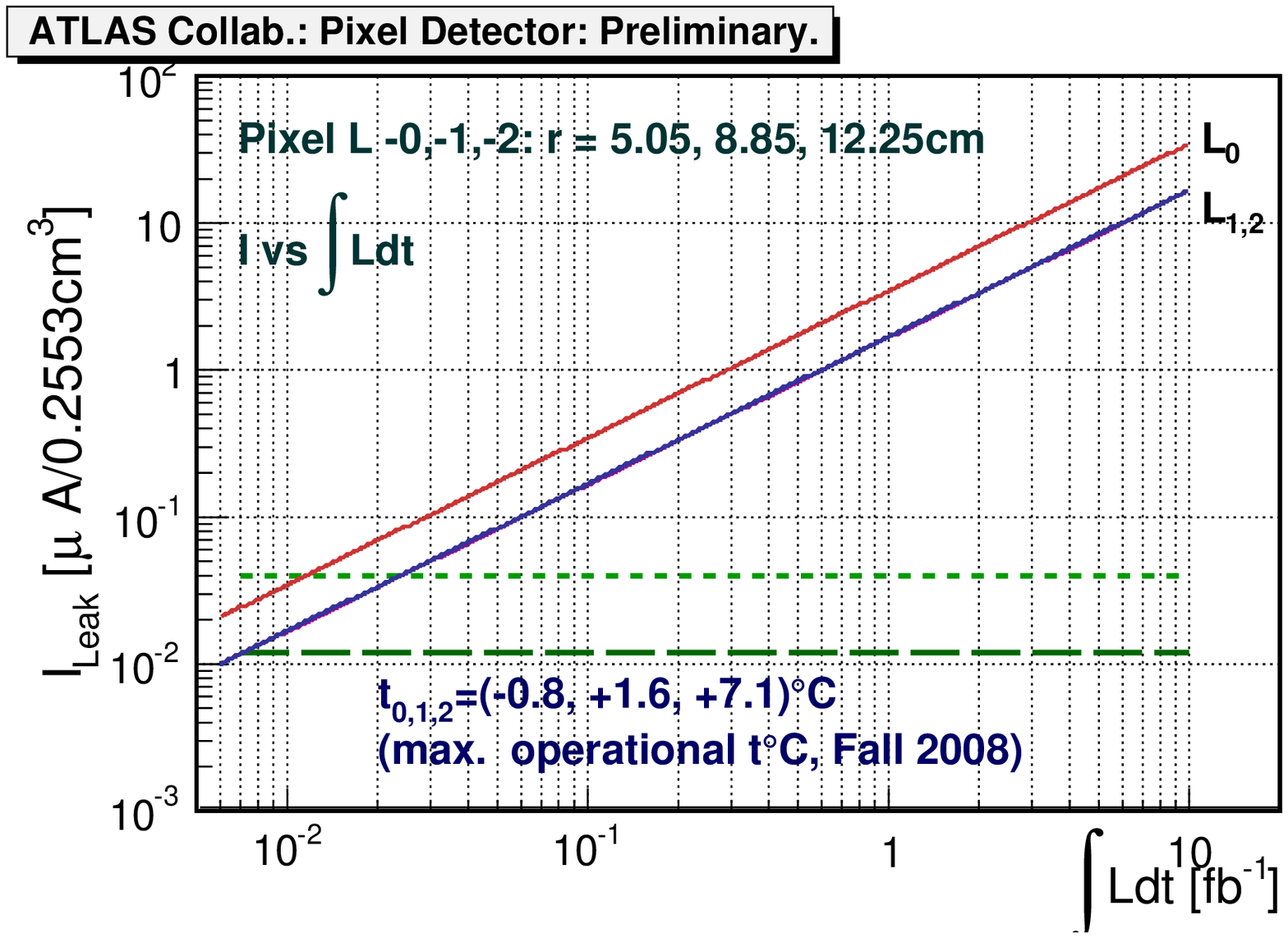}\\
      \includegraphics[width=85mm]{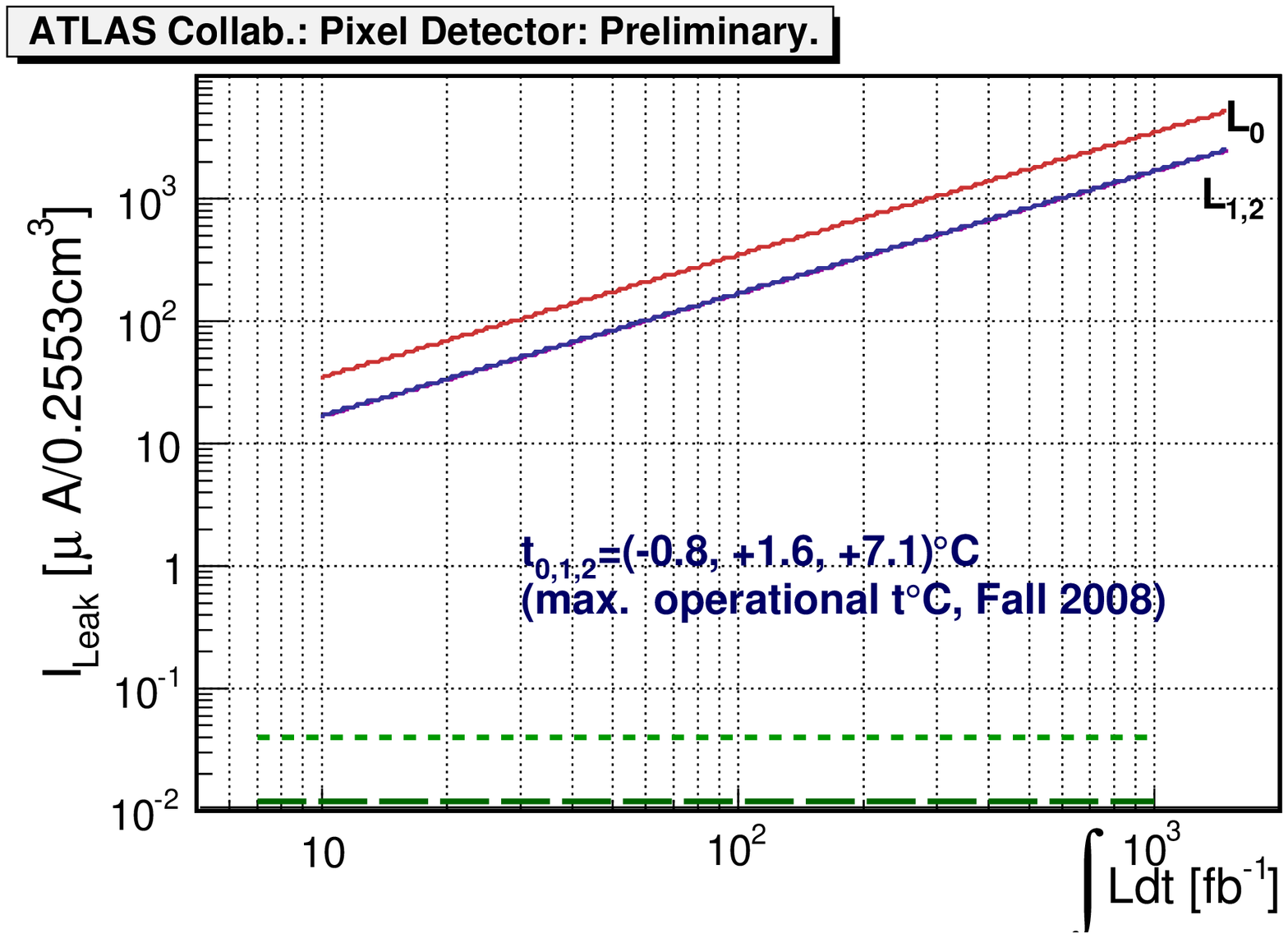}
    \end{center}
  \end{figure}

  The most recent temperature data allow us to apply the temperature
  corrections under realistic conditions.  Shown in
  Fig.~\ref{fig:pix-maxtC-current} are {\it the leakage current
  readings at maximum temperatures reached during the cosmic run of
  Fall 2008}.
  Table~\ref{table:current-maxtC} shows fluences and leakage current
  values for several luminosities and for {\it Layer-0}, {\it Layer-1}
  and {\it Layer-2} at the maximum temperatures specified in the
  table.
\begin{table}[htbp]
  \begin{center}
  \caption{Currents predicted by MC \textbf{at the maximum temperatures 
           recorded during the Fall 2008} cosmic run.}
\begin{tabular}{|r|r|r|}
\hline
 \multicolumn{3}{|l|}{\bf ATLAS Pixel Layer-0: \(t_{max} = -0.8\degc\)}\\
\hline
    \(\IntL,\invfb\) & \(\flue, \cm^{-2}\) & \(I_{Leak},\mu{A}\)\\
\hline
     0.100&  2.428193e+11 &  3.325354e-01 \\
     1.000&  2.428193e+12 &  3.325354e+00 \\ 
    10.000&  2.428193e+13 &  3.325354e+01 \\ 
   100.000&  2.428193e+14 &  3.325354e+02 \\ 
  1000.000&  2.428193e+15 &  3.325354e+03 \\ 
  1400.000&  3.399471e+15 &  4.655496e+03 \\ 
  1500.000&  3.642290e+15 &  4.988031e+03 \\
\hline
 \multicolumn{3}{|l|}{\bf ATLAS Pixel Layer-1: \(t_{max} = +1.6\degc\)}\\
\hline
    \(\IntL,\invfb\) & \(\flue, \cm^{-2}\) & \(I_{Leak},\mu{A}\)\\
\hline
     0.100&  9.119346e+10 & 1.597900e-01 \\ 
     1.000&  9.119346e+11 & 1.597900e+00 \\ 
    10.000&  9.119346e+12 & 1.597900e+01 \\ 
   100.000&  9.119346e+13 & 1.597900e+02 \\ 
  1000.000&  9.119346e+14 & 1.597900e+03 \\ 
  1400.000&  1.276708e+15 & 2.237059e+03 \\ 
  1500.000&  1.367902e+15 & 2.396849e+03 \\ 
\hline
 \multicolumn{3}{|l|}{\bf ATLAS Pixel Layer-2: \(t_{max} = +7.1\degc\)}\\
\hline
    \(\IntL,\invfb\) & \(\flue, \cm^{-2}\) & \(I_{Leak},\mu{A}\)\\
 \hline
     0.100&  5.326114e+10 &  1.616612e-01 \\ 
     1.000&  5.326114e+11 &  1.616612e+00 \\ 
    10.000&  5.326114e+12 &  1.616612e+01 \\ 
   100.000&  5.326114e+13 &  1.616612e+02 \\ 
  1000.000&  5.326114e+14 &  1.616612e+03 \\ 
  1400.000&  7.456560e+14 &  2.263257e+03 \\ 
  1500.000&  7.989171e+14 &  2.424918e+03 \\ 
\hline
\end{tabular}
    \label{table:current-maxtC}
  \end{center}
\end{table}
\par
\begin{figure}[htbp]
    \begin{center}
      \caption{The expected currents \textbf{at minimal operational
               temperature values reached in Fall 2008 Cosmic Run by
               the Pixel Detector} barrel {\it Layer-0}, {\it Layer-1}
               and {\it Layer-2} versus the integrated luminosity in
               two ranges \( \IntL\in(10\invpb,\,10\invfb) \) (upper
               plot) and \( \IntL\in(10\invfb,\,1500\invfb) \) (lower
               plot).  {\bf Two levels of sensitivity of
               the proposed Current Measurement Board, \( 0.01\mkamp\)
               (optimistic) and \( 0.04\mkamp\) (realistic), are shown
               as dashed green lines }. The different
               slopes due to different fluences through {\it Layer-0},
               {\it Layer-1}, and {\it Layer-2} are seen in
               logarithm-scaled coordinates as the offsets between
               lines. \label{fig:pix-mintC-current}  } 
      \includegraphics[width=85mm]{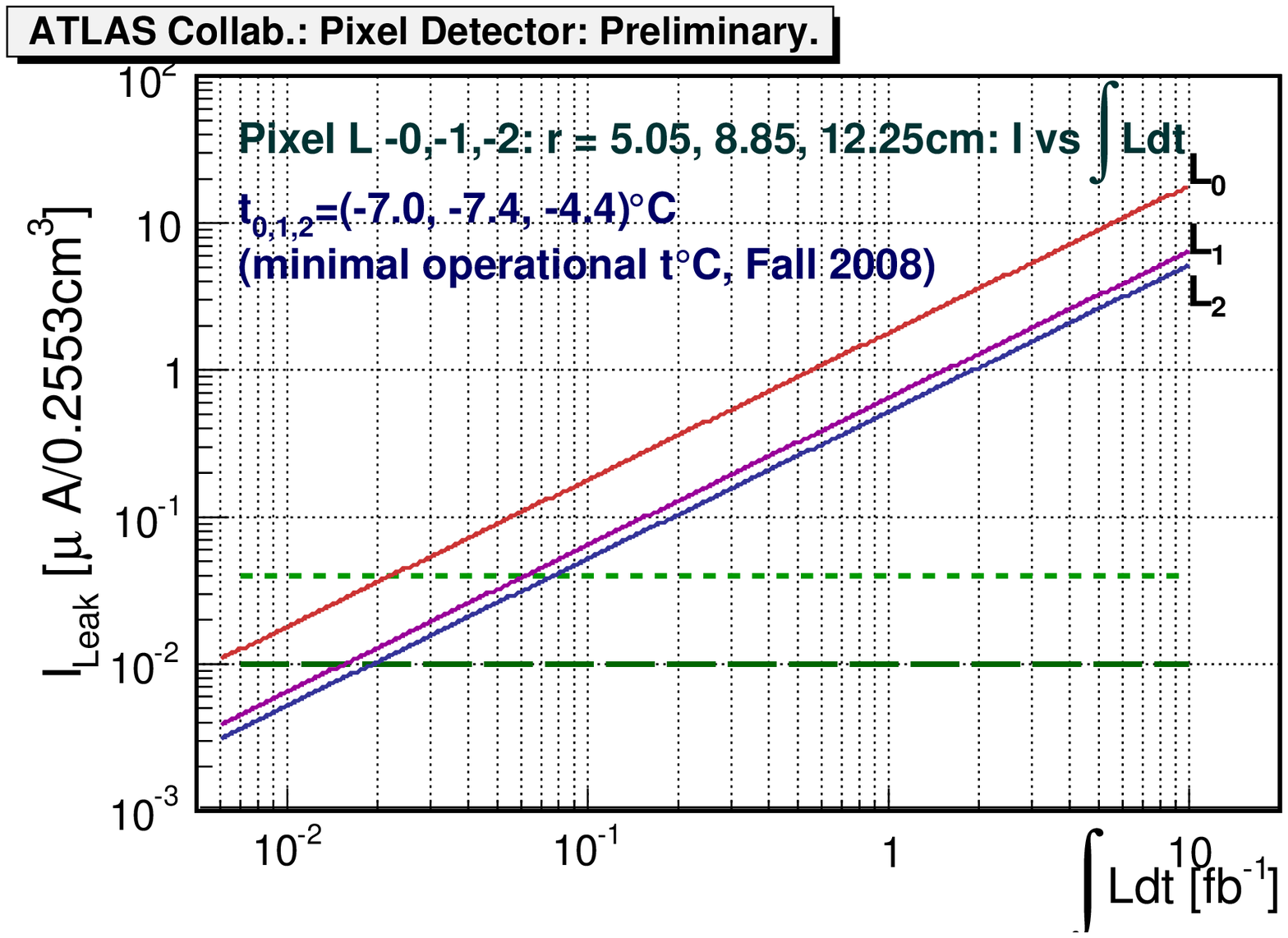}\\
      \includegraphics[width=85mm]{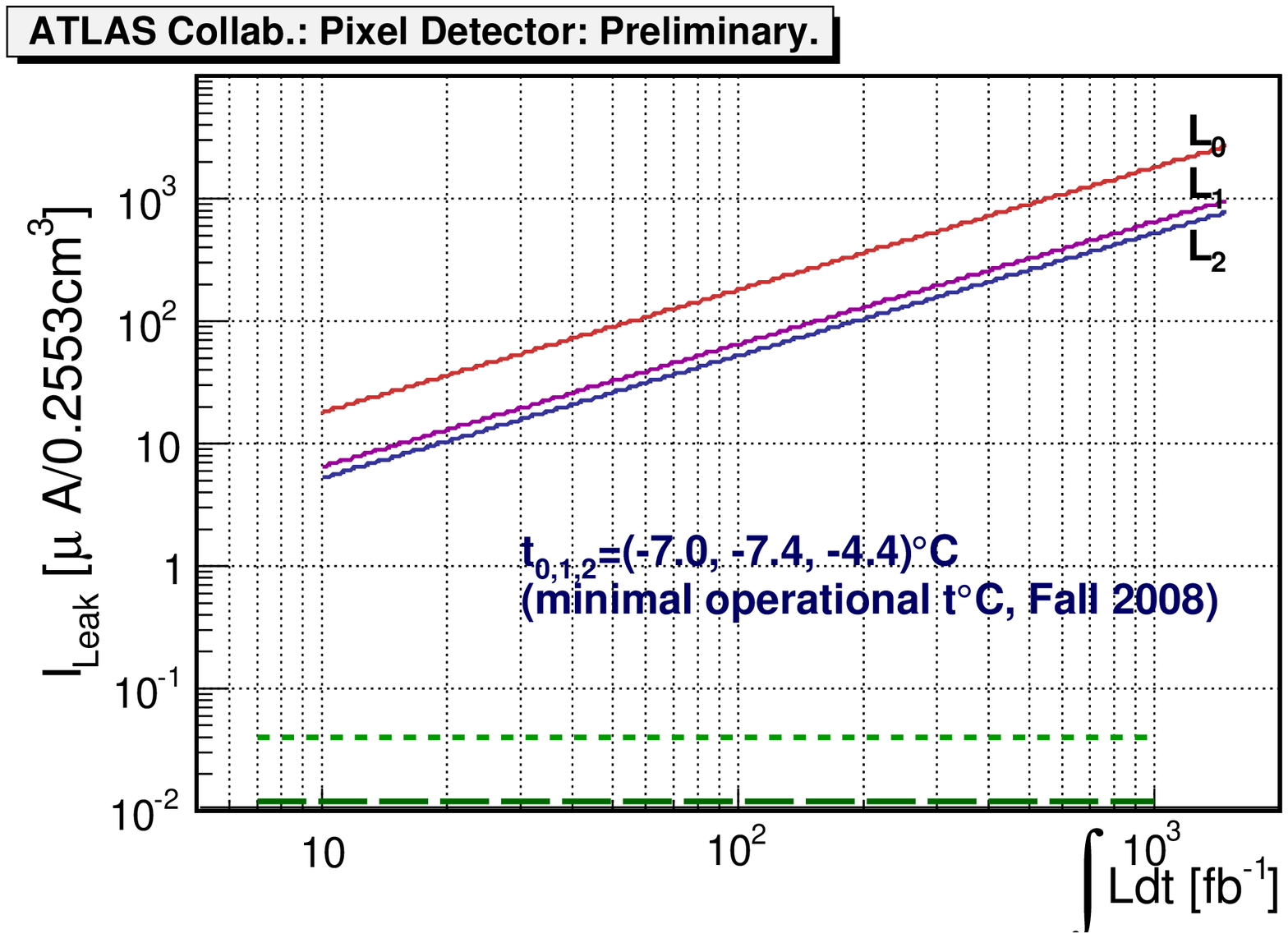}
    \end{center}
  \end{figure}

  Fig.~\ref{fig:pix-mintC-current} shows the predicted leakage
  currents up to \(1500\invfb \) for {\it the minimum temperatures
  reached during the Fall 2008 cosmic run}. { We expect that these
  are close to realistic operational conditions. The plots also show
  two levels of sensitivity of the current measurement board: case 1,
  \( \mathbf{0.01\mkamp} \) (optimistic expectation) and case 2, 
  \( \mathbf{0.04\mkamp} \) (realistic expectation).} The sensitivity
  level is a crucial technical specification for the HVPP4 Current
  Measurement Board.
  Table~\ref{table:current-mintC} shows fluences and leakage current
  values for several luminosities and for {\it Layer-0}, {\it Layer-1}
  and {\it Layer-2} at {\it the minimum temperatures} specified in the
  table. Based on the values presented in the table, one can evaluate
  the ratio between the minimum (at \( \IntL\sim100\invpb \) ) and
  maximum (at \( \IntL\sim1500\invpb \) ) expected currents to be 
  \( \sim{0.5}\cdot{10^{5}} \).  The required dynamic range of leakage
  currents to be processed is technically challenging.
\par
%
%
%
%
\begin{table}[htbp]
\begin{center}
\caption{Currents predicted at 
         \textbf{the minimum temperatures recorded during Fall 2008} 
         cosmic run. }
\begin{tabular}{|r|r|r|}
  \hline
  \multicolumn{3}{|l|}{\bf ATLAS Pixel Layer-0: \(t_{min} = -7.0\degc\)}\\
  \hline
    \(\IntL,\invfb\) & \(\flue, \cm^{-2}\) & \(I_{Leak},\mu{A}\)\\
  \hline
     0.100&  2.428193e+11 &  1.724755e-01   \\
    1.000&   2.428193e+12 &  1.724755e+00   \\ 
   10.000&   2.428193e+13 &  1.724755e+01   \\ 
   100.000&  2.428193e+14 &  1.724755e+02   \\ 
  1000.000&  2.428193e+15 &  1.724755e+03   \\ 
  1400.000&  3.399471e+15 &  2.414657e+03   \\ 
  1500.000&  3.642290e+15 &  2.587133e+03   \\
  \hline
  \multicolumn{3}{|l|}{\bf ATLAS Pixel Layer-1: \(t_{min} = -7.4\degc\)}\\
  \hline
    \(\IntL,\invfb\) & \(\flue, \cm^{-2}\) & \(I_{Leak},\mu{A}\)\\
  \hline
     0.100&  9.119346e+10 &  6.202588e-02   \\ 
     1.000&  9.119346e+11 &  6.202588e-01   \\ 
    10.000&  9.119346e+12 &  6.202588e+00   \\ 
   100.000&  9.119346e+13 &  6.202588e+01   \\ 
  1000.000&  9.119346e+14 &  6.202588e+02   \\ 
  1400.000&  1.276708e+15 &  8.683623e+02   \\ 
  1500.000&  1.367902e+15 &  9.303882e+02   \\ 
 \hline
 \multicolumn{3}{|l|}{\bf ATLAS Pixel Layer-2: \(t_{min} = -4.4\degc\)}\\
 \hline
    \(\IntL,\invfb\) & \(\flue, \cm^{-2}\) & \(I_{Leak},\mu{A}\)\\
 \hline
     0.100&  5.326114e+10 &   4.999913e-02  \\ 
     1.000&  5.326114e+11 &   4.999913e-01  \\ 
    10.000&  5.326114e+12 &   4.999913e+00  \\ 
   100.000&  5.326114e+13 &   4.999913e+01  \\ 
  1000.000&  5.326114e+14 &   4.999913e+02  \\ 
  1400.000&  7.456560e+14 &   6.999878e+02  \\ 
  1500.000&  7.989171e+14 &   7.499870e+02  \\ 
  \hline
  \end{tabular}
  \label{table:current-mintC}
\end{center}
\end{table}
\subsection{ Disk Layers: Preliminary Estimates of the Slopes of 
            \( \Delta{I}/V \) versus \( \IntL \) }
  In previous sections our estimates and considerations are focused on
  the barrel layers.  Following a description in~\cite{jinst-mech-serv}
  eight disk sectors are mounted on a 312 mm diameter carbon composite
  disk support ring, forming a disk. There are three disks in each of
  the two end-caps. Three modules are mounted on each side of the sector,
  with the long dimension of the module in the radial direction. The
  three modules on the back side of the sector are rotated 
  \( 7.5\degrees \) with respect to the modules on the front side, making
  the overlapping to provide a full acceptance in \(\theta\) (or
  pseudo-rapidity \( \eta \)).  Each disk has on back and front sides 
  \( 2\times24\,=\,48 \) modules.  Each end cap comprises 
  \( 3\times48\,=\,144 \) modules with a total of 
  \( 2\times144\,=\,288 \) modules for both end caps.
\par 
  The radius of the module centers is approximately 
  \( R_{\rm{disk\,module}}^{center}\approx119\mm\) . 
  The inner radius of the active area of the pixel 
  modules \( R_{\rm{disk\,module}}^{inner}\approx89\mm\).  
  Please see the details in~\cite{jinst-mech-serv}.
\par 
  We follow the general conjecture made in~\cite{atlas:rad-bgr} about 
  weak \(z\)-dependence of the fluence in the Pixel Detector area
  especially in the barrel region. We extrapolate this approach to the 
  endcap area taking the ``worst case scenario''. 
  Thenceforth to estimate the fluence through disk modules we should
 integrate the dependence in Eq.~\ref{eq:flue-poly} over radial area of
  \( (88.88,\,149.6)\mm \) and take an average. 
  Please see the Eq.~\ref{eq:flue-disk}.  
 \begin{eqnarray}
   < \flue^{disk} > & \!\propto & \!\!\int_{R_{out}}^{R_{inn}} \!\!\!{d\phi}\cdot{r}{dr}\cdot\flue{(r)} 
\label{eq:flue-disk}
 \end{eqnarray}
   Using the same numbers from Ian Dawson's recent
   update~\cite{atlas:rad-bgr} on fluence in ATLAS Pixel Detector area
   including disks (see also a discussion in Section~\ref{subsec:mcsim}), 
   \(R_{inn} = 8.88\cm,\,R_{out} = 14.96\cm,\,\) and 
   \( a_{-2}=4.93\cdot10^{+16}, \, a_{-1}=0.25\cdot10^{+16} \), 
   the averaged over disk module fluence per 1000\invfb can be calculated to be
   \[ < \flue^{disk.sens.} > \times1000\invfb \approx 5.64\cdot10^{+14}\,\mathrm{cm^{-2}} \] 
  Below the Table~\ref{table:current-disk-mintC} shows the fluences and leakage currents
  values for several luminosities and for {\it Disk Layer}
  at {\it the minimum and maximum Fall 2008 temperatures} specified in the
  table. Again here as for the barrel case we assume that all modules are drawing 
  the currents of the same value.
\begin{table}[htbp]
\begin{center}
  \caption{Currents in disk layer predicted at \textbf{the minimum and
           maximum temperatures recorded during Fall 2008} cosmic
           run. }
  \begin{tabular}{|r|r|r|}
  \hline
  \multicolumn{3}{|l|}{\bf ATLAS Pixel Disk Layers: \(t_{min} = -7.3\degc\)}\\
  \hline
    \(\IntL,\invfb\) & \(\flue, \cm^{-2}\) & \(I_{Leak},\mu{A}\) \\
  \hline
     0.100&  5.645341e+10 &  3.881622e-02 \\
     1.000&  5.645341e+11 &  3.881622e-01 \\ 
    10.000&  5.645341e+12 &  3.881622e+00 \\ 
   100.000&  5.645341e+13 &  3.881622e+01 \\ 
  1000.000&  5.645341e+14 &  3.881622e+02 \\ 
  1400.000&  7.903477e+14 &  5.434271e+02 \\ 
  1500.000&  8.468011e+14 &  5.822433e+02 \\
  \hline
   \multicolumn{3}{|l|}{\bf ATLAS Pixel Disk Layers: \(t_{max} = -3.4\degc\)}\\
  \hline
    \(\IntL,\invfb\) & \(\flue, \cm^{-2}\) & \(I_{Leak},\mu{A}\) \\
  \hline
     0.100&  5.645341e+10 &  5.891448e-02  \\
     1.000&  5.645341e+11 &  5.891448e-01  \\ 
    10.000&  5.645341e+12 &  5.891448e+00  \\ 
   100.000&  5.645341e+13 &  5.891448e+01  \\ 
  1000.000&  5.645341e+14 &  5.891448e+02  \\ 
  1400.000&  7.903477e+14 &  8.248027e+02  \\ 
  1500.000&  8.468011e+14 &  8.837172e+02  \\
  \hline
\end{tabular}
  \label{table:current-disk-mintC}
\end{center}
\end{table}
\par 
  The temperatures at disk layer areas have been held lower than for
  outer barrel layers {\it Layer-1} and {\it Layer-2}. The fluences at
  disk area shown in Table~\ref{table:current-disk-mintC} are similar to
  the ones at barrel {\it Layer-2}, see Table~\ref{table:current-mintC}.
  Thus the leakage currents in disk layer at its minimal temperature are
  similar to the ones of barrel {\it Layer-2}, see
  Table~\ref{table:current-mintC}, but somewhat lower at its maximum
  temperature reached during Fall 2008. That said we conjecture that the
  specified range of the currents to be sensed by the Current Measurement
  Board is in compliance with the currents to be drawn by the disk
  modules given the temperatures observed during Fall 2008 cosmic runs.
%
%

%% file: hardware/board.tex
\section{Current Measurement Board} 
\label{sec:board}
\par 
  The present HVPP4 System serves as a fan-out point for the bias
  voltages delivered by Type II boards from Iseg power supplies to 1744
  pixel modules. The current measurement function of HVPP4 system is
  technically implemented by Current Measurement Board mounted on every
  Type II fan-out board.  
\par 
  The analog current measurements are further digitized by the ATLAS
  standard 64-channel ELMB board~\cite{elmb-manual}~\cite{elmb-firmware} and sent
  via CAN bus to DCS database (see also~\cite{Aad:2008zz}).
  The ADC serving every of the ELMB channels can be configured for
  a full-scale measurement of the voltage coming from Current
  Measurement Board in the next 5 ranges~\cite{elmb-firmware}:
  \begin{itemize}
    \item \( V_{input}\in(0.,25)\mvolt \), \( V_{input}\in(0.,100)\mvolt \), ...
    \item ...\( V_{input}\in(0.,1)\volt \), \( V_{input}\in(0.,2.5)\volt \), ... 
    \item \( V_{input}\in(0.,5)\volt \).
    \item the \(16\) bits of ADC provides the resolution 
         of \( (0, 65535) \)~\cite{elmb-manual}
         for an every of above specified ranges. 
  \end{itemize}
\par
\begin{figure}[htbp]
\begin{center}
\caption{The response of a pre-prototype current measurement board. The
          upper plot shows the response of the board to the calibrated
          current supplied by Keithley power source. The middle plot
          includes also the current drawn by biased ATLAS tile and chip
          sensors. The lowest plot is a result of the tests made at SR1
          ATLAS pit area using Keithley power source feeding currents
          into the measurement board, ELMB digital board processing
          analog output of the measurement board and the PVSS software
          reading out and formatting the digital data. \label{fig:board-prepro} }
%
%
\includegraphics[width=80mm]
{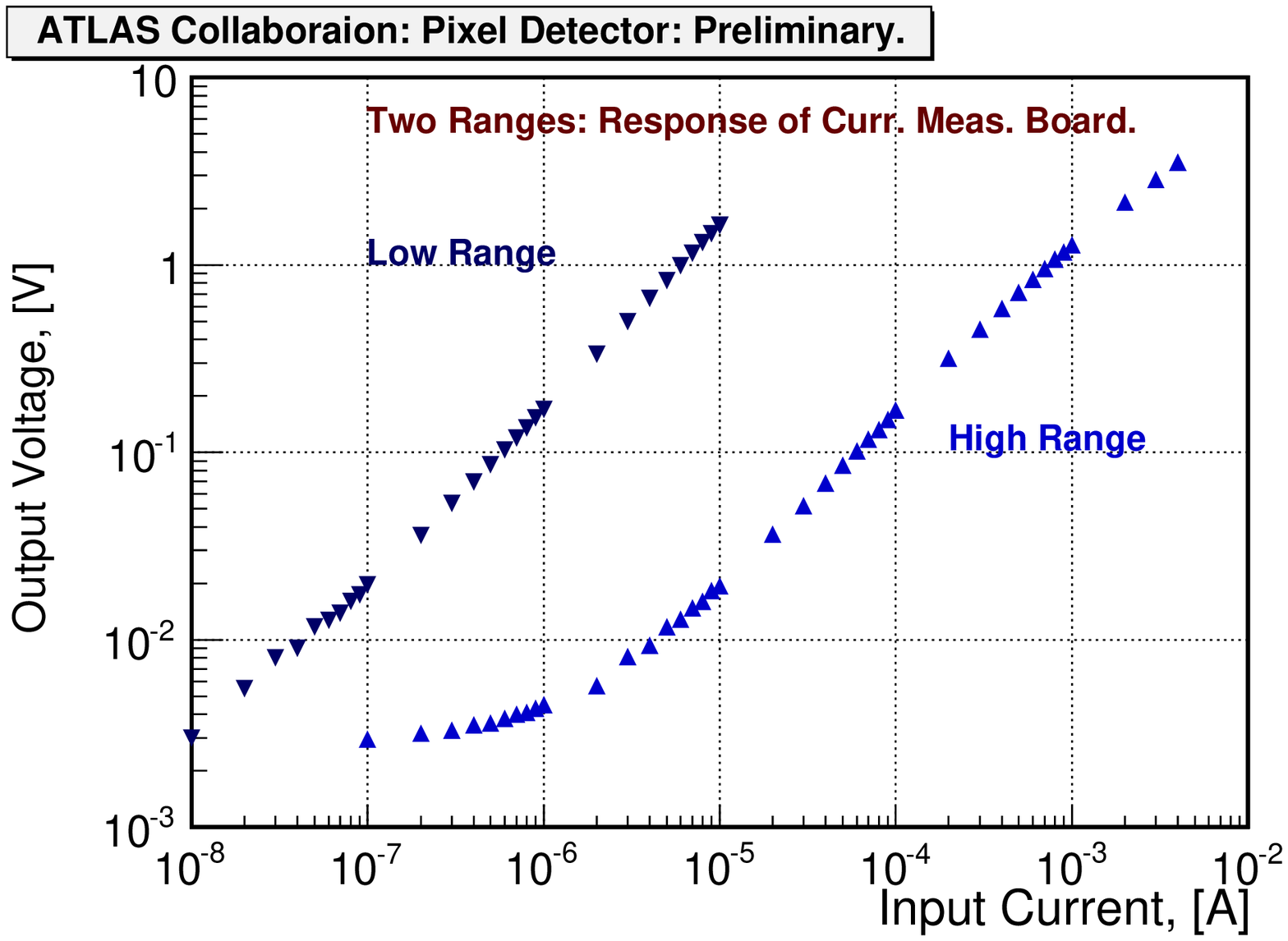}
\includegraphics[width=80mm]
{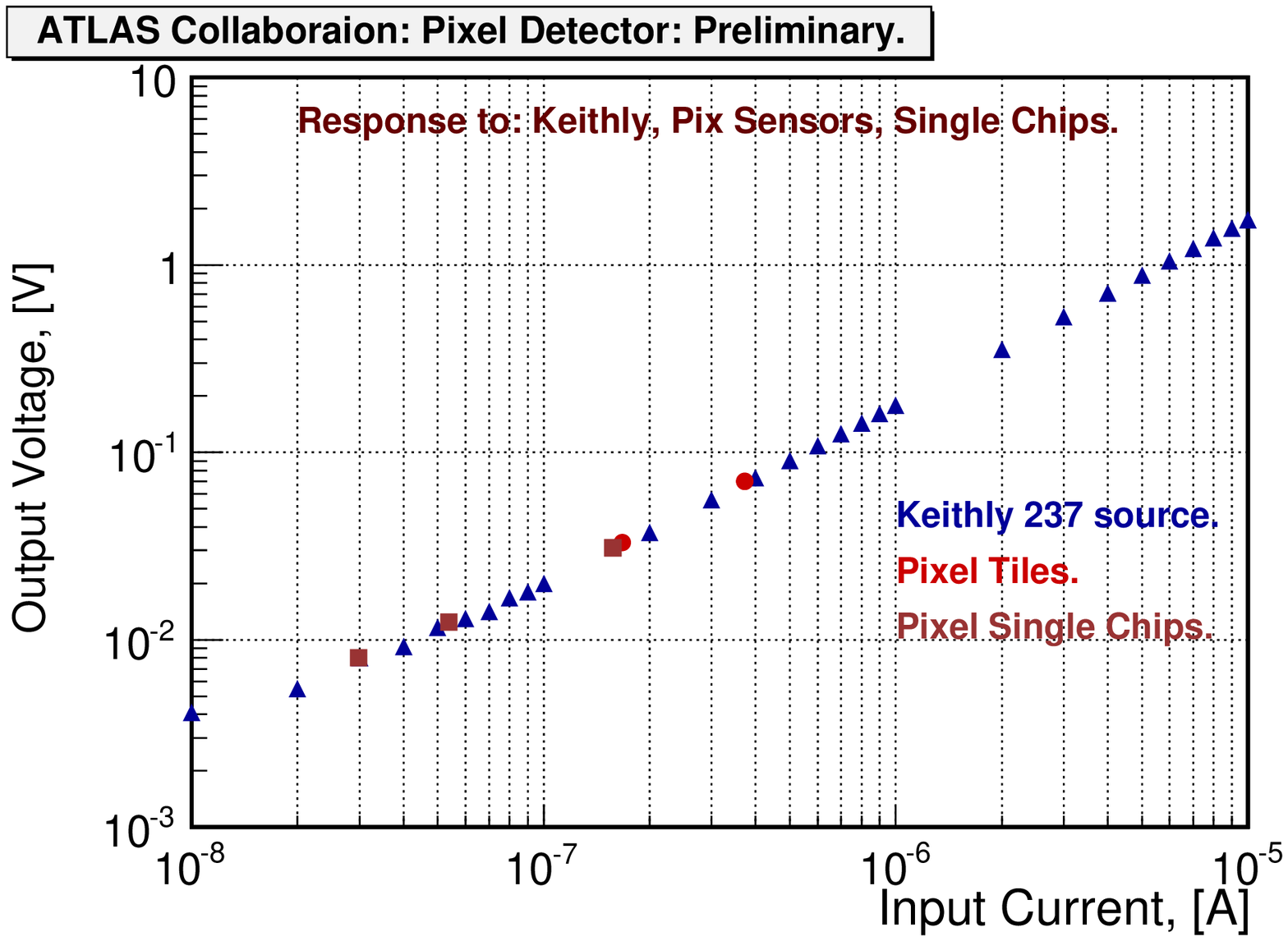} 
\includegraphics[width=80mm]
{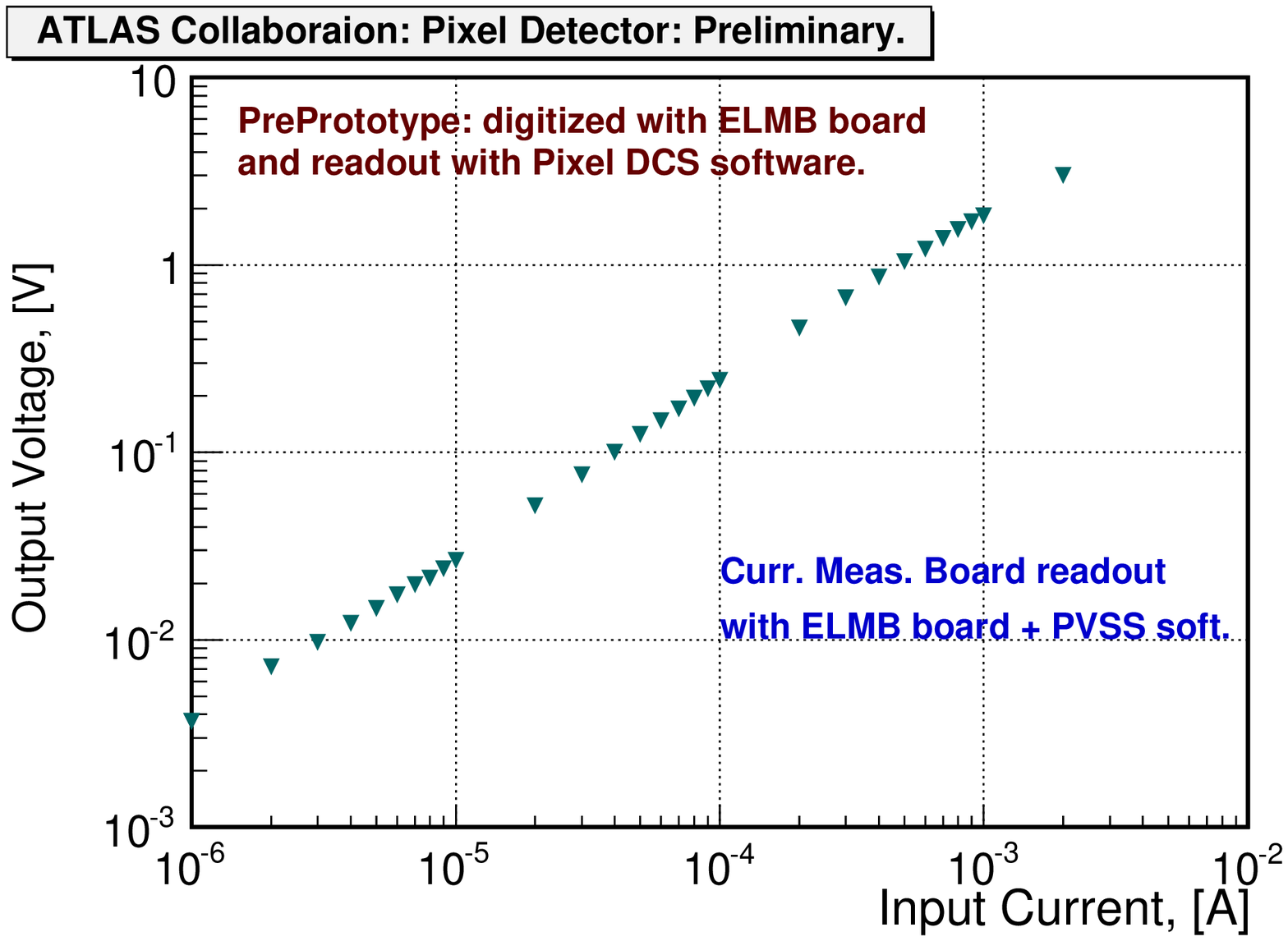}  
%
%
%
\end{center}
\end{figure}

  The specification for a range of currents to be measured with the
  board comes from our estimates of currents for the expected integrated
  luminosities to be delivered by LHC, see Fig.~\ref{fig:pix-maxtC-current}
  and Fig.~\ref{fig:pix-mintC-current}.
  \begin{itemize}
    \item \( (0.04\mkamp\,,\,2\mamp) \) with a dynamical range of 
          \( \sim{0.5}\times{10^{5}} \)
    \item the output voltage of the board should lie within 
          \( (0.\,,\,5)\rm{V_{DC}} \) 
          to comply the digital board ELMB specifications outlined 
          above.
    \item the circuit of the Current Measurement Board is a current to
          frequency converter which in turn is optically coupled to a
          frequency to voltage converter
    \begin{itemize}
      \item the pairs of channels are isolated from each other and
    \end{itemize}
    \item the board is a multi-layer PCB holding 13 current measurement
          circuits and providing the current measurements for 13
          channels (pixel modules)
    \item the pairs of channels are isolated from each other and
          from the pixel module readout system
  \end{itemize}
  The configuration will consist of several VME crates:
  \begin{itemize}
    \item One VME crate filled with 9 Type II boards
    \begin{itemize}
      \item Current Measurement Board mounted on every Type II board, 9
            boards per VME crate
      \item 13 channels per Current Measurement Board: \(9\times13 \)
            channels per crate
      \item 2 ELMB boards to digitize and send data over CAN bus to ATLAS
            DCS database
   \end{itemize}
   \item In total the HVPP4 system consists of 16 VME crates to serve
         \( 16\times9\times13\,=\,1872 \) channels well enough for 1744
         modules
  \end{itemize}
  Recently the pre-prototype of the Current Measurement Board was laid out and produced.
  The board has been tested with calibrated current source from Keithley, with real ATLAS
  tile and chip sensors biased to the appropriate voltage at \( 20\degc \). 
  The responses of the pre-prototype board are shown at plots, see Fig.~\ref{fig:board-prepro}.
  The nice linearity has been observed.
%
%
%

%% file: summary/summ.tex
\section{ Summary }
\label{sec:summ}  
  We have described the principles of radiation damage monitoring using
  the current measurements to be provided by the circuits of the HVPP4
  system.  The dependence of the leakage current with respect to the
  integrated luminosity at several temperature scenarios has been
  presented. Based on the analysis we have evaluated the sensitivity
  specifications for the Current Measurement Board to be a crucial
  subsystem of HVPP4. The pre-prototype of the Current Measurement Board
  has been developed, produced and tested with real ATLAS
  sensors and at SR1 area in ATLAS pit. A regular linear behavior of the
  response has been obtained with real ATLAS sensors at \( 20\degc \)
  and with a calibrated current source.
%
%
\begin{acknowledgments}
  The authors are grateful to Dr. Maurice Garcia-Sciveres (LBNL) and
  Rusty Boyd (University of Oklahoma) for their contributions and
  continuous support of the project.
\end{acknowledgments}

%
%